# Abrupt Change Detection in Power System Fault Analysis using Adaptive Whitening Filter and Wavelet Transform

Abhisek Ukil

*(Corresponding Author)*

*Zethushof 209, Park Street 620, Arcadia, Pretoria, 0083, South Africa*

*Tshwane University of Technology*

*Tel: +27 (0)72 736 9557*

*Fax: +27 (0)12 460 7440*

*E-mail:* abhiukil@yahoo.com

Rastko Živanović

*P.O. Box 8484, Pretoria, 0001, South Africa*

*Tshwane University of Technology*

*E-mail:* zivanovr@yahoo.com

**Abstract**

This paper describes the application of the adaptive whitening filter and the wavelet transform used to detect the abrupt changes in the signals recorded during disturbances in the electrical power network in South Africa. Main focus has been to estimate exactly the time-instants of the changes in the signal model parameters during the pre-fault condition and following events like initiation of fault, circuit-breaker opening, auto-reclosure of the circuit-breakers. The key idea is to decompose the fault signals, de-noised using the adaptive whitening filter, into effective detailed and smoothed version using the multiresolution signal decomposition technique based on discrete wavelet transform. Then we apply the threshold method on the decomposed signals to estimate the change time-instants, segmenting the fault signals into the event-specific sections for further signal processing and analysis. This paper presents application on the recorded signals in the power transmission network of South Africa.

# 1    Introduction

Detection of abrupt changes in the signal characteristics has a significant role to play in failure detection and isolation (FDI) systems; one such domains, power system fault analysis is the focus of this paper. In this paper, we propose the use of adaptive whitening filter and wavelet transform, particularly the dyadic-orthonormal wavelet transform for estimating the time-instants of the abrupt changes in the power system fault signals recorded during the disturbances in the electrical power transmission network of South Africa.

In this paper, adaptive whitening filter based on the adjusted Fourier filter [1] is used to pre-filter the original fault signal. Wavelet transform is used to transform the pre-filtered fault signal into the finer wavelet scales, followed by a progressive search for the largest wavelet coefficients on that scale [2]. Large wavelet coefficients that are co-located in time across different scales provide estimates of the changes in the signal parameter. The change time-instants can be estimated by the time-instants when the wavelet coefficients exceed a given threshold (which is equal to the *'universal threshold'* of Donoho and Johnstone [3] to a first order of approximation).

The remainder of this paper is organized as follows. In section-2, power system fault analysis as application domain is discussed. In section-3, adaptive whitening filter is reviewed. Wavelet transform is reviewed in section-4. Section-5 discusses the pre-filtering operation using the adaptive whitening filter and the signal decomposition using the wavelet transform. Utilization of the threshold method for segmentation is explained in section-6. Practical application results are presented in section-7, and conclusions are given in section-8.

# 2    Power System Fault Analysis

We consider the power system fault analysis as our application domain, focusing on the automatic disturbance recognition and analysis for the ESKOM power transmission network in South Africa. Presently, 98% of the transmission lines are equipped with the Digital Fault Recorders (DFRs) on the feeder bays, with an additional few installed on the Static Var

Compensators (SVCs) and 95% of these are remotely accessible via a X.25 communication system [4].

Majority of digital fault recorders in ESKOM are from Siemens: 'SIMEAS R' and 'OSCILLOSTORE P531' recorders. 'SIMEAS R' comes with 16-bit resolution and a 12.8 kHz maximum scanning frequency per channel. There are two types of central unit: one with 8 analogue and 16 binary channels, and one with 32 analogue and 64 binary channels [4]. 'OSCILLOSTORE P531' comes with 8- and 12-bit resolution and a 5 kHz maximum scanning frequency. It has 31 Data acquisition units (DAUs) per central unit, i.e., 124 analogue or 992 binary data acquisition channels [4]. Every feeder uses 3 Data acquisition unit (DAU): 2 x ADAU (analogue data acquisition units used for 4 x voltage and 4 x current signals), and 1 x BDAU (binary data acquisition unit used for up to 32 digital signals) [4]. The 32 Binary values are either stored as a '0' or a '1' and indicates the status of a contact e.g., breaker auxiliary contact. Analogue values indicate the magnitude of an analogue signal (voltage or current) measured at a specific point in time.

The DFRs and the associated settings are applied for the purpose of protection performance and disturbance analysis. The DFRs trigger due to reasons like power network fault conditions, protection operations, breaker operation and the like. Following IEEE Comtrade standard [5], the DFR recordings are provided as input to the analysis software which uses Discrete Fourier Analysis and Superimposed current quantities [4].

The purpose of this study is to augment the existing fault analysis system with more robust and accurate algorithms and techniques to make it fully automated. So, we would first apply the abrupt change detection algorithms to segment the fault recordings into different segments, namely pre-fault segment, after initiation of fault, after circuit-breaker opening, after auto-reclosure of the circuit-breakers. Then we would construct the appropriate feature vectors for the different segments; finally the pattern-matching algorithm would be applied using those feature vectors to accomplish the fault recognition and analysis tasks. In the scope of this paper, we focus on the first task i.e., segmentation of the fault recordings by detecting the abrupt changes in the characteristics of the fault recordings, following the architecture as shown in Figure 1.

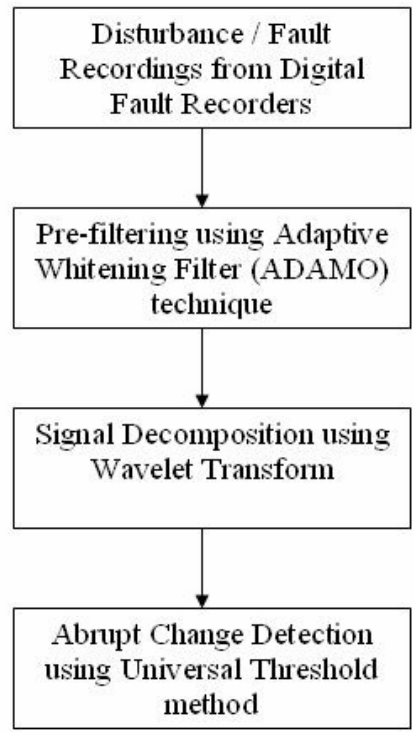

Figure 1. Architecture for Abrupt Change Detection

Following the comparative study of the different techniques [6] for the abrupt change detection, we propose the use of the adaptive whitening filter and the wavelet transform. Wavelet transform is particularly suitable for the power system disturbance and fault signals which may not be periodic and may contain both sinusoidal and impulse components. Also, for the power system fault analysis, time-frequency resolution is needed which states another reason for using the wavelet transform because it provides a local representation (both in time and frequency) of a given signal unlike the Fourier transform which provides a global representation of a signal. In particular, the ability of the wavelets to focus on short intervals for high-frequency components and long intervals for low-frequency components improves the decomposition of the fault signals into finer and detailed scales, facilitating further effective signal processing and analysis.

## 3     Adaptive Whitening Filter

The optimal Linear Prediction Error (LPE) filter for transient monitoring would perfectly de-correlate the signal, leaving only white noise (whitening filter) [1]. The finite impulse response (FIR) whitening filter can be defined by

$$A(z^{-1}) = 1 - z^{-C}, \tag{1}$$

where *A* indicates the finite impulse response of the whitening filter, *z* indicates the Z-transform and *C* is the rounded number of samples per cycle at the nominal network frequency.

$$C = round\left(\frac{f_s}{f_{fund}}\right), \tag{2}$$

where $f_s$ is the sampling frequency and $f_{fund}$ is the fundamental frequency; (*C* = 50 at 50 Hz for a sampling frequency of 2.5 kHz).

This filter compares the current signal value $x_k$ to the value approximately one cycle before, $x_{k-C}$. The zeroes of this filter are the harmonics of the fundamental signal, up to the Nyquist frequency, they are evenly spaced on the unit cycle [1]. This whitening filter is called the Fourier filter [7].

When the true fundamental frequency is close to $f_s / C$, but not exactly equal to this quantity, the zeroes of the whitening filter move away from the true harmonics, however, it is still possible to force a filter zero at the estimated fundamental frequency with some more computation. This leads to the Adjusted Fourier filter [7] in case of a non-integer frequency ratio.

Using the same definition of *C* as before, let

$$A(z^{-1}) = 1 - \alpha z^{-C+1} - \beta z^{-C}, \tag{3}$$

where $\alpha$ and $\beta$, functions of the network frequency, are given by

$$z^C - \alpha z - \beta = 0 \qquad \text{for} \qquad z = e^{\pm j\omega_0 T_s}, \tag{4}$$

where $T_s$ is the sampling period and $\omega_0$ is the network pulsation. We obtain the following formulae [1] for computing the coefficients $\alpha$ and $\beta$:

$$\alpha = \frac{\sin(\omega_0 T_s C)}{\sin(\omega_0 T_s)}, \tag{5.1}$$

$$\beta = \cos(\omega_0 T_s C) - \alpha \cos(\omega_0 T_s). \tag{5.2}$$

Adaptive whitening filter is based on the Adjusted Fourier filter [7] with the fact that the output of the filter must be minimum when its coefficients are well adapted. A Least Mean Square (LMS) estimate of $\alpha$ and $\beta$ is carried out to minimize the output, with the constraint of exactly filtering the dc component of the filter [1]. Derivation of the complete recursive equations of the Adaptive Whitening filter following the LMS method can be found in [1] and [7].

This filter has the advantage of perfectly extracting the main frequency component of the signal, strongly attenuating its harmonics and the dc component. Characteristics of the adaptive whitening filter in terms of its frequency response are shown in Figure 2.

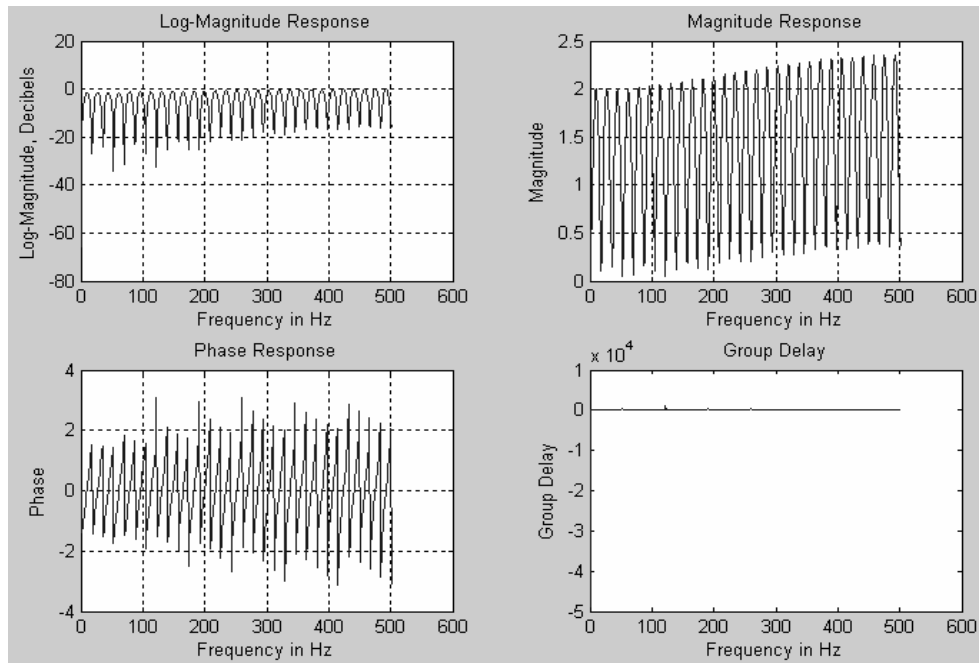

Figure 2. Frequency Response plots of the Adaptive Whitening Filter

## 4    Wavelet Transform

The Wavelet Transform (WT) is a mathematical tool, like Fourier transform for signal analysis. A wavelet is an oscillatory waveform of effectively limited duration that has an average value of zero. Fourier analysis consists of breaking up a signal into sine waves of various frequencies. Similarly, wavelet analysis is the breaking up of a signal into shifted and scaled versions of the original (or mother) wavelet. In Figure 3, we show the basis functions for the Fourier transform (Sine wave) and Wavelet transform (*db10*: Daubechies 10 mother wavelet [8]).

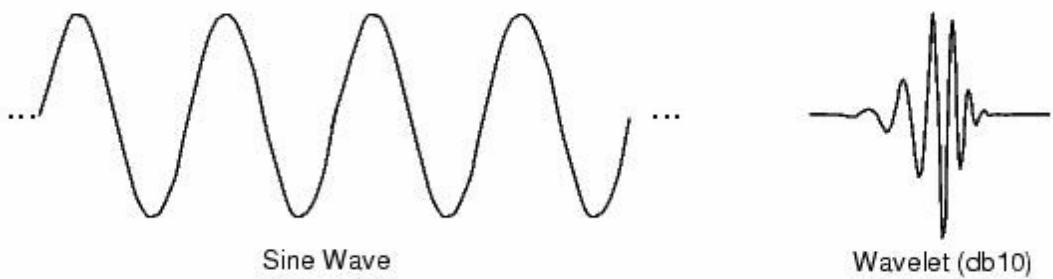

Figure 3.  Basis Functions for Fourier Transform & Wavelet Transform

Fourier analysis does not provide good results for the *non- stationary* signals, i.e., where the signal parameters change over the time unlike the *stationary* signals. This is because in transforming the complete signal to the frequency domain, the time information gets lost in the Fourier analysis. This deficiency in the Fourier analysis can be overcome to some extent by analyzing a small section of the signal at a time - a technique called *windowing* the signal, first proposed by Dennis Gabor. This leads to an analysis technique called Short-Time Fourier Transform (STFT). But the drawback in STFT is that the size of the time-window is same for all frequencies. Wavelet analysis overcomes this deficiency by allowing a windowing technique with variable-sized regions, i.e.*,* wavelet analysis allows the use of the long time intervals where we want more precise low-frequency information, and the shorter regions where we want high-frequency information. In Figure  4, we show the time-domain (Shannon), frequency-domain (Fourier), STFT (Gabor) and wavelet views of signal analysis.

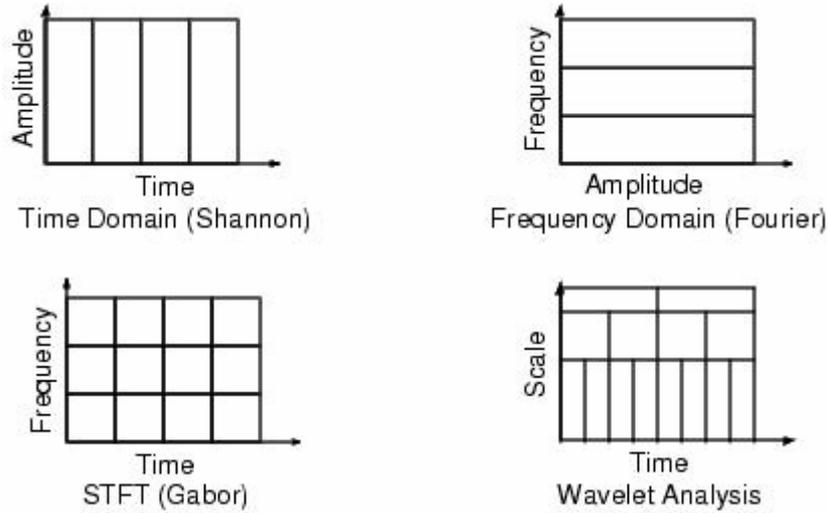

Figure 4. Time, Frequency, STFT, Wavelet views of signal analysis

While detail mathematical descriptions of WT can be referred to in [8], [9], a brief mathematical summary of WT is provided in the following sections in relation to the application domain within the scope of this paper.

*4.1    Continuous Wavelet Transform*

The Continuous Wavelet Transform (CWT) is defined as the sum over all time of the signal multiplied by the scaled and shifted versions of the wavelet function $\psi$. The CWT of a signal *x*(*t*) is defined as

$$CWT(a,b) = \int_{-\infty}^{\infty} x(t)\psi^*_{a,b}(t)\, dt, \qquad (6)$$

where

$$\psi_{a,b}(t) = |a|^{-1/2}\psi((t-b)/a). \qquad (7)$$

$\psi(t)$ is the *mother* wavelet, the asterisk in (6) denotes a complex conjugate, and $a,b \in R, a \neq 0$, (*R* is a real continuous number system) are the *scaling* and *shifting* parameters respectively.

$|a|^{-1/2}$ is the normalization value of $\psi_{a,b}(t)$ so that if $\psi(t)$ has a unit length, then its scaled version $\psi_{a,b}(t)$ also has a unit length.

*4.2    Discrete Wavelet Transform*

Instead of continuous scaling and shifting, the mother wavelet maybe scaled and shifted discretely by choosing $a = a_0^m, b = na_0^m b_0, t = kT$ in (6) & (7), where $T = 1.0$ and $k, m, n \in Z$, ($Z$ is the set of positive integers). Then, the Discrete Wavelet Transform (DWT) is given by

$$DWT(m,n) = a_0^{-m/2}\left(\sum x[k]\psi^*[(k - na_0^m b_0)/a_0^m]\right). \quad (8)$$

By careful selection of $a_0$ and $b_0$, the family of scaled and shifted mother wavelets constitutes an orthonormal basis. An orthonormal basis is a basis that consists of a set of vectors **S** such that $\mathbf{u}.\mathbf{v} = 0$ (here '.' indicates the dot product) for each distinct pair of $\mathbf{u}, \mathbf{v} \in \mathbf{S}$. We can choose $a_0 = 2$ and $b_0 = 1$ to constitute the orthonormal basis to have the WT called a *dyadic-orthonormal* WT. The implications of the dyadic-orthonormal WT is that due to the orthonormal properties there will be no information redundancy among the decomposed signals. Also, with this choice of $a_0$ and $b_0$, there exists a novel algorithm, known as *multiresolution signal decomposition* [10] technique, to decompose a signal into scales with different time and frequency resolution.

*4.3    Multiresolution Signal Decomposition and Quadrature Mirror Filter*

The Multiresolution Signal Decomposition (MSD) [10] technique decomposes a given signal into its detailed and smoothed versions. Let *x*[*n*] be a discrete-time signal, then MSD technique decomposes the signal in the form of WT coefficients at scale 1 into $c_1[n]$ and $d_1[n]$, where $c_1[n]$ is the smoothed version of the original signal, and $d_1[n]$ is the detailed version of the original signal *x*[*n*]. They are defined as

$$c_1[n] = \sum_k h[k-2n]\,x[k], \tag{9}$$

$$d_1[n] = \sum_k g[k-2n]\,x[k], \tag{10}$$

where $h[n]$ and $g[n]$ are the associated filter coefficients that decompose $x[n]$ into $c_1[n]$ and $d_1[n]$ respectively. Downsampling is done in the process of decomposition so that the resulting $c_1[n]$ and $d_1[n]$ are each $n/2$ point signals. Thus, for the original $n$ point signal $x[n]$, after the decomposition we have $n$ point signal together with $c_1[n]$ and $d_1[n]$, not $2n$ point.

The next higher scale decomposition will be based on $c_1[n]$. Thus, the decomposition process can be iterated, with successive approximations being decomposed in turn, so that the original signal is broken down into many lower resolution components. This is called the *wavelet decomposition tree* [9].

MSD technique can be realized with the cascaded *Quadrature Mirror Filter* (QMF) [11] banks. A QMF pair consists of two finite impulse response filters, one being a low-pass filter (LPF) and the other a high-pass filter (HPF). The QMF pair divides the input signal into low-frequency and high-frequency components at the dividing point of halfway between zero hertz and half the data sampling frequency. The output of the low-pass filter is the smoothed version of the input signal and used as the next QMF pair's input. The output of the high-pass filter is the detailed version of the original signal. Figure 5 shows the MSD technique and the QMF pairs.

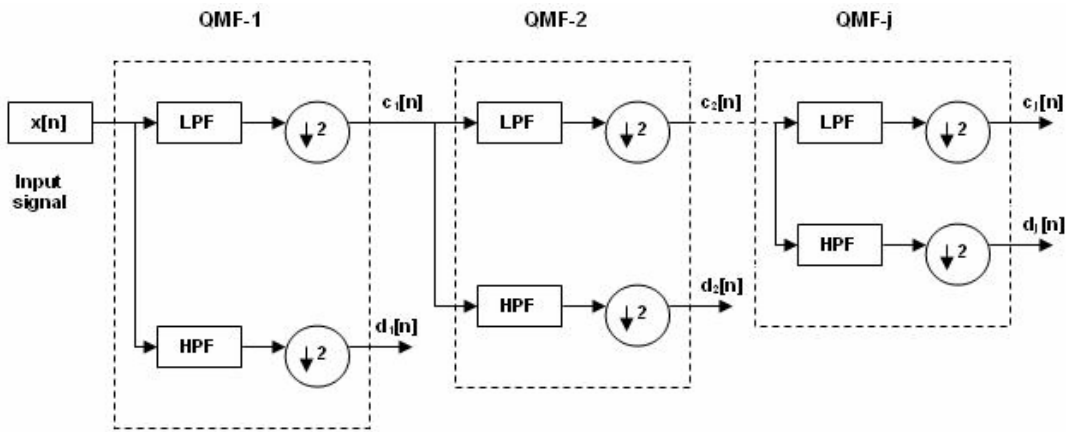

Figure 5. Multiresolution Signal Decomposition

## 5      Pre-filtering and Signal Decomposition

In this section, we discuss the pre-filtering and signal decomposition operations on the fault signals, utilizing the adaptive whitening filter and the wavelet transform respectively.

### 5.1      Pre-filtering

The fault signals from the DFRs are first filtered with the adaptive whitening filter discussed in section-3. A sampling frequency of 2.5 kHz, same as that of the DFRs [4], is used for the adaptive whitening filter with primary focus on the 50 Hz component. So, we set the network pulsation frequency at 51 Hz, at close proximity of the 50 Hz. This choice of the network pulsation frequency is the optimal choice for this application following several tests with different fault signals.

The pre-filtering operation extracts the main frequency component to be used for the signal decomposition, strongly attenuating its harmonics and the dc component. This operation increases the precision and sensitivity of the next operations, namely signal decomposition and abrupt change detection. Besides, this pre-filtering operation particularly contributes to distinguish a fault from a transient recovery, a short-term swing and the like, which otherwise are very difficult to estimate using only the wavelet transform.

### 5.2      Signal Decomposition

On the pre-filtered and de-noised fault signals, we apply the multiresolution signal decomposition technique and quadrature mirror filter banks to decompose the signals into localized and detailed representation in the form of wavelet coefficients. Daubechies 4 [8] wavelet is used as the mother wavelet, i.e., the filters $h[n]$ and $g[n]$ as in (9) & (10) are chosen with four coefficients and calculated as in [8]. Among many other choices of the mother wavelets, e.g., Coiflets, Meyer wavelet, Gaussian wavelet, Mexican hat wavelet, Morlet wavelet and the like [8], Daubechies 4 wavelet is chosen because Daubechies wavelets are compactly supported [8] wavelets with extremal phase and highest number of vanishing moments for a given support width [8], also the associated scaling filters are minimum-phase filters [8]. So, from the point of views of fast implementation and varying patterns of the fault signals, Daubechies wavelets appear to be the optimal choice for the mother wavelet for this specific application. The order 4 for the Daubechies wavelet has been chosen based on the optimal performance characteristics conforming to the varying patterns of the fault signals.

For the Daubechies 4 wavelet, the scaling function $\phi(x)$ has the form

$$\phi(x) = c_0 \phi(2x) + c_1 \phi(2x-1) + c_2 \phi(2x-2) + c_3 \phi(2x-3), \tag{11}$$

where

$$c_0 = (1+\sqrt{3})/4, \tag{12.1}$$

$$c_1 = (3+\sqrt{3})/4, \tag{12.2}$$

$$c_2 = (3-\sqrt{3})/4, \tag{12.3}$$

$$c_3 = (1-\sqrt{3})/4. \tag{12.4}$$

It is not possible in general to solve directly for $\phi(x)$; the approach is to solve for $\phi(x)$ iteratively until $\phi_j(x)$ is very nearly equal to $\phi_{j-1}(x)$, where

$$\phi_j(x) = c_0 \phi_{j-1}(2x) + c_1 \phi_{j-1}(2x-1) + c_2 \phi_{j-1}(2x-2) + c_3 \phi_{j-1}(2x-3). \tag{13}$$

The Daubechies 4 wavelet function $\psi(x)$ for the four-coefficient scaling function is given by

$$\psi(x) = -c_3 \phi(2x) + c_2 \phi(2x-1) - c_1 \phi(2x-2) + c_0 \phi(2x-3). \tag{14}$$

The Daubechies 4 scaling function and wavelet function are shown in Figure 6 (a) & (b) respectively.

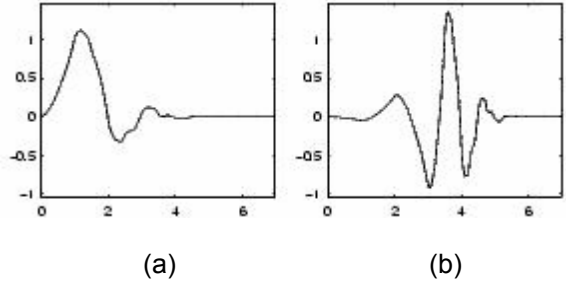

Figure 6.  Daubechies 4: (a) Scaling Function, (b) Wavelet Function

After transforming the original fault signal using the mother wavelet described above, we obtain the smoothed and detailed versions, i.e., $c_1[n]$ and $d_1[n]$ [see (9), (10)] respectively. Signal $d_1[n]$ can be considered to be the difference between the original signal $x[n]$ and $c_1[n]$, and called the wavelet transform coefficient at scale one. We will use $d_1[n]$ for threshold checking to estimate the change time-instants described in the following section.

# 6    Application of Threshold Method

We will use the threshold method on the wavelet transform coefficients of the pre-filtered fault signal to detect the jumps and sharp cusps [12] in order to estimate the time-instants of the abrupt changes. Mathematically we say that signal $x(t)$ has a sharp $\alpha$-cusp at $t$ if for $0 \leq \alpha < 1$,

$$|x(t + \Delta t) - x(t)| \geq K|\Delta t|^{\alpha}, \qquad (15)$$

as $\Delta t \to 0$ for some constant $K \geq 0$. We can consider it to be a jump if $\alpha = 0$. In practice, we can consider a cusp to be an abrupt change of the level of the trend over a small time period.

As discussed in the previous section, after transforming the original fault signal using the wavelet transform, we will search progressively across the finer wavelet scales for the largest wavelet coefficients on that scale [2]. As wavelet coefficients are changes of averages, so a coefficient of large magnitude implies a large change in the original signal. Large wavelet coefficients that are co-located in time across different scales provide estimates of the cusp points [2] hence time-instants of the abrupt changes. The change time-instants can be estimated by the instants when the wavelet coefficients exceed a given threshold which is equal to the '*universal threshold*' of Donoho and Johnstone [3] to a first order of approximation.

The universal threshold *T* is given by

$$T = \sigma\sqrt{2\log_e n} \, , \qquad (16)$$

where $\sigma$ is the median absolute deviation of the wavelet coefficients, divided by 0.6725 [3] and *n* is the number of samples of the wavelet coefficients. Instead of standard deviation median absolute deviation is used because median is hardly influenced by a small fraction of extreme values [12].

After determining the time-instants when the wavelet coefficients of the fault signal exceed the threshold, we mark them using the unit impulses, indicating the abrupt change time-instants.

## 7    Application Results

In this section, we present the practical application results of the abrupt change detection in power system fault analysis developed according to the above discussed signal pre-filtering and decomposition using the adaptive whitening filter and the WT and then applying the threshold method on the detailed version of the fault signal, followed by heuristic smoothing filtering operation. MATLAB® with Wavelet toolbox [13] has been used for implementing the application. The whole procedure detects the abrupt change time-instants thus segments the fault signal. We are interested in the change time-instants to be indicated as unit impulses.

First the original fault signal is normalized using its mean value. The normalized fault signal is then filtered using the adaptive whitening filter. The filtered signal is then transformed into the smoothed and detailed version using the WT and then the threshold method is applied on the detailed version to determine the change time-instants. Then smoothing filter operations are applied on this segmented model to perform sequentially the following smoothing operations:

- It removes confusing multiple close-spikes and combines them into single unit impulse.

- It removes any unwanted glitches which can otherwise result in false positives for the abrupt changes.

- The segments in the power system fault analysis signals are during the pre-fault condition and following events like fault initiation, circuit-breaker opening and reclosing. These events are predefined and so are the number of segments. So, any bigger number of segmentation possibly indicates transients, power swings and the like. Estimation of the number of segment(s) is also performed and checked to distinguish the fault from the transients, power swings etc.

- Based on the modeling of the segments, analysis is done for estimating the event-critical change instants, rejecting others.

In Figure 7, we show the result for the fault signal obtained from the DFRs in the electrical power network of ESKOM, South Africa during a phase to ground fault. The fault signal is sampled at a sampling frequency of 2.5 kHz.

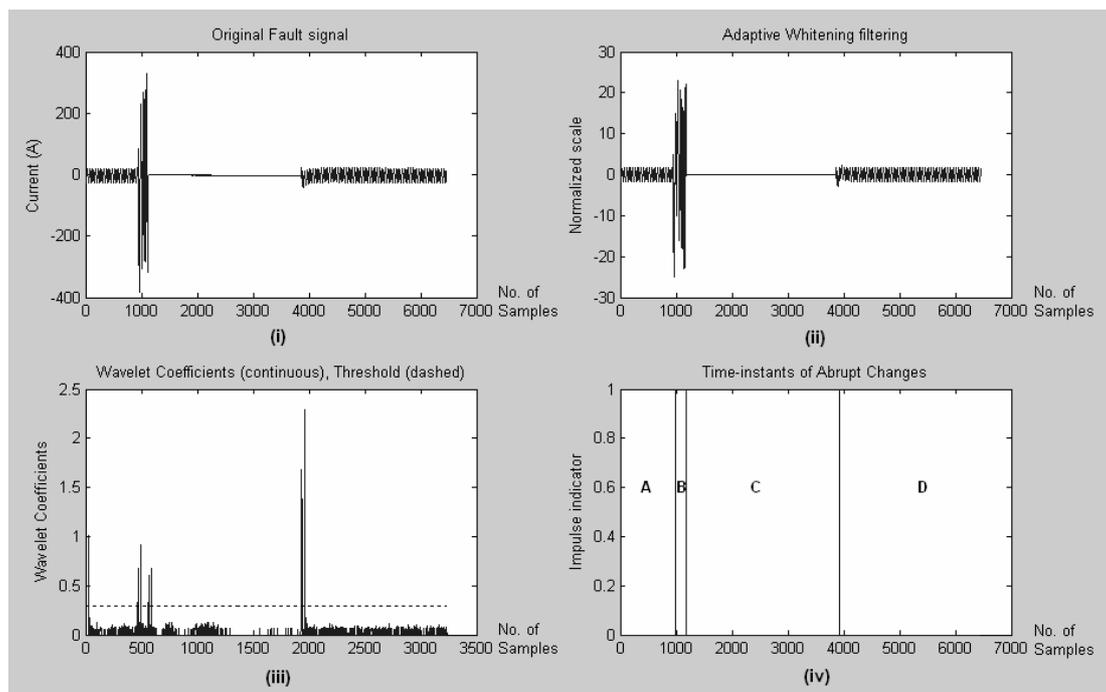

Figure 7. Segmentation of the RED-Phase Current signal

In Figure 7, the original DFR recording for the current during the fault in the RED-Phase is shown in the plot (i), pre-filtering of the fault signal using the adaptive whitening filter is shown in the plot (ii), wavelet coefficients for the filtered fault signal (in continuous line) and the universal threshold (in dashed line) are shown in the plot (iii) and the change time-instants computed using the threshold checking followed by smoothing filtering is shown in the plot (iv). It is to be noted that only the high-pass filter output of the QMF pair is shown for the wavelet coefficients, so the wavelet coefficients in the plot (iii) indicate half of the total number of samples of the original signal. In the plot (iv), the time-instants of the abrupt changes in the signal characteristics indicated by the impulse indicators show the different signal segments owing to the different events during the fault, e.g., segment A indicates the pre-fault section and the fault inception, segment B indicates the fault, segment C indicates opening of the circuit-breaker, segment D indicates auto-reclosing of the circuit breaker and system restore.

Figure 8 shows another result for the fault signal obtained from the DFRs in the electrical power network of ESKOM, South Africa during a phase to phase fault. The fault signal is sampled at a sampling frequency of 2.5 kHz.

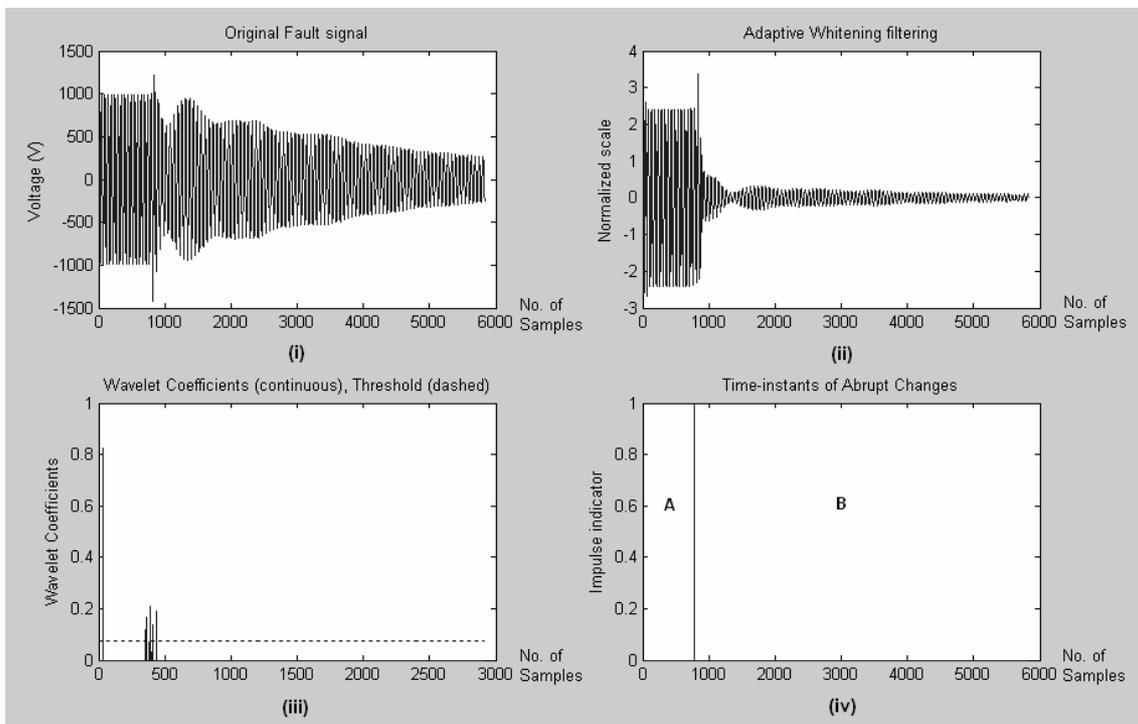

Figure 8. Segmentation of the RED-Phase Voltage signal

In Figure 8, the original DFR recording for the voltage in the RED-Phase during the fault involving the RED- and BLUE- Phases is shown in the plot (i), pre-filtering of the fault signal using the adaptive whitening filter is shown in the plot (ii), wavelet coefficients for the filtered fault signal (in continuous line) and the universal threshold (in dashed line) are shown in the plot (iii) and the change time-instants computed using the threshold checking followed by smoothing filtering is shown in the plot (iv). It is to be noted that only the high-pass filter output of the QMF pair is shown for the wavelet coefficients, so the wavelet coefficients in the plot (iii) indicate half of the total number of samples of the original signal.

In Figure 8, we show the case where the fault signal follows a resistive decay, which does not produce the sharp transitions among the different segments due to the different events during the fault as in the case shown in Figure 7. As in these cases we cannot determine the different sharp segments, we rather focus on correctly estimating the time-instant of the fault-inception for further signal processing and analysis based on it. In Figure 8, this is shown in plot (iv), where the impulse indicator shows the fault-inception time-instant, thus segmenting the fault signal into two segments, pre-fault and post-fault segment indicated by the segment A and B respectively. The application of the adaptive whitening filter for pre-filtering the fault signal is of particular importance for these kind of signals, as shown in the plot (ii) of the Figure 8 that the filtered fault signal clearly suppresses the resistive decay part and highlights the fault-inception time-instant, which improves the accuracy of detection of the fault-inception time-instant using the subsequent wavelet decomposition and threshold operations.

Following the discussion of the applied algorithms and the application results, the following comments can be cited.

- The intended application is not meant for real-time analysis, so computation time is not a critical factor. However, the proposed algorithm for the abrupt change detection and signal segmentation took an average computation time of 0.501 seconds. An Intel® Celeron® 1.9 GHz computer was used for all the application tests using MATLAB® [13]. It is to be noted that the complete automatic disturbance recognition and analysis tasks have to be performed within five minutes of the acquiring of the fault signals, abrupt change detection and segmentation being the first step.

- The proposed algorithm using the adaptive whitening filter and the wavelet transform is considerably faster and more robust compared to the traditional peak value detection and superimposed current quantities algorithms [6]. Also, the automatic segmentation based on abrupt change detection using the wavelet transform and threshold method facilitates further signal processing and analysis in the subsequent stages of automatic disturbance recognition and analysis, focusing on the different segments and helping to determine quickly parameters like duration of the fault etc directly from the abrupt change detection based segmentation itself, which cannot be done using the traditional peak value detection and superimposed current quantities algorithms [6].

- Instead of Fourier transform, Wavelet transform is particularly suitable for the power system disturbance and fault signals which may not be periodic and may contain both sinusoidal and impulse components.

- Wavelet coefficients are greatly adaptive to the fault signal pattern variations.

- For the power system fault analysis, time-frequency resolution is needed which can be achieved by the wavelet transform because it provides a local representation (both in time and frequency) of a given signal, and *Multiresolution Signal Decomposition* allows short intervals for high-frequency components and long intervals for low-frequency components. This is not possible with the traditional Fourier transform which provides a global representation of a signal.

- Use of the adaptive whitening filter improves the accuracy of the abrupt change detection, particularly when the fault signals consist of resistive decay and the like, not showing sharp-changing segments. Use of this pre-filtering operation also contributes to distinguish a fault from a transient recovery, a short-term power swing and the like, which otherwise are very difficult to estimate using only the wavelet transform.

## 8      Conclusion

We have presented in this paper the application of the adaptive whitening filter and the wavelet transform for detecting the abrupt changes in the signals recorded during disturbances in the transmission network of South Africa. Power system disturbance and fault signals may not be periodic and may contain both sinusoidal and impulse components. So, we first filter the fault signals using the adaptive whitening filter, focusing on the main frequency component to be used for signal decomposition, strongly attenuating its harmonics and the dc component. Then we propose the use of wavelet transform, particularly the dyadic-orthonormal wavelet transform to decompose the filtered fault signal into smoothed and detailed version in terms of wavelet coefficients using the multiresolution signal decomposition technique. Then we make a progressive search on that wavelet scale for the largest wavelet coefficients. The change time-instants can be estimated by the time-instants when the wavelet coefficients exceed a given threshold (which is equal to the *'universal threshold'* of Donoho and Johnstone [3] to a first order of approximation). This is followed by smoothing operation. The results obtained from the MATLAB® implementation are quite good. So, the use of the adaptive whitening filter to pre-filter the fault signal and the dyadic-orthonormal wavelet transform to transform the filtered fault signal into the smoothed and detailed version, followed by threshold checking are quite effective in detecting the abrupt changes in the signals originating from power system faults and segmenting them into the event-specific sections, facilitating and improving the accuracy of the automatic disturbance recognition and analysis task.

**Acknowledgements**

This work was supported in part by the National Research Foundation (NRF), South Africa. All real fault signal recordings were kindly provided by ESKOM, South Africa.

**Vitae**

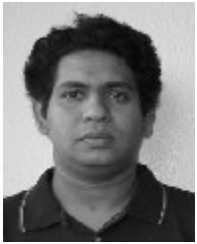

**Abhisek Ukil** received the B.E. degree in electrical engineering from the Jadavpur University, Calcutta, India, in 2000 and the M.Sc. degree in electronic systems and engineering management from the University of Applied Sciences, South Westphalia, Soest, Germany, and Bolton Institute, UK, in 2004.

Currently, he is pursuing the D.Tech. degree at Department of Mathematical Technology, Tshwane University of Technology, Pretoria, South Africa.

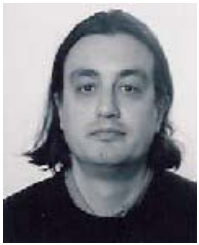

**Rastko Živanović** received the Dipl.Ing. and M.Sc. degrees from the University of Belgrade, Belgrade, Serbia, in 1987 and 1991, and the Ph.D. degree from the University of Cape Town, Cape Town, South Africa, in 1997. Currently, he is Professor with the Faculty of Engineering at Tshwane University of Technology, Pretoria, South Africa, where he has also been Lecturer and Senior Lecturer since 1992.

His research interests include power system protection and control.

**Figure Captions**

Figure 1. Architecture for Abrupt Change Detection

Figure 2. Frequency Response plots of the Adaptive Whitening Filter

Figure 3. Basis Functions for Fourier Transform & Wavelet Transform

Figure 4. Time, Frequency, STFT, Wavelet views of signal analysis

Figure 5. Multiresolution Signal Decomposition

Figure 6. Daubechies 4: (a) Scaling Function, (b) Wavelet Function

Figure 7. Segmentation of the RED-Phase Current signal

Figure 8. Segmentation of the RED-Phase Voltage signal